\documentclass[conference]{IEEEtran}
\usepackage{lipsum}
\usepackage{amsthm}
\usepackage{framed} 

\newtheoremstyle{mystyle}%                % Name
  {}%                                     % Space above
  {}%                                     % Space below
  {\itshape}%                             % Body font
  {}%                                     % Indent amount
  {\bfseries}%                            % Theorem head font
  { }%                                    % Punctuation after theorem head
  {.5em}%                                    % Space after theorem head, ' ', or \newline
  {}%                                     % Theorem head spec (can be left empty, meaning `normal')

\theoremstyle{mystyle}

\usepackage{ifpdf}
\usepackage{cite}

\usepackage[colorlinks=false,linkcolor=red]{hyperref}
\usepackage{amsmath,dsfont,bbm,epsfig,amssymb,amsfonts,amstext,verbatim,amsopn,cite,subfigure,multirow,multicol,lipsum,xfrac}
\usepackage{mathtools}
\usepackage{perpage}
\usepackage{url}
\usepackage{amsfonts}
\usepackage[font={small}]{caption}
\usepackage{stmaryrd}
\usepackage{psfrag}	
\usepackage{multirow}
\usepackage[process=auto]{pstool}
\usepackage{etoolbox}
\usepackage{algpseudocode}
\usepackage{pifont}
\usepackage[utf8]{inputenc}
\usepackage[T1]{fontenc}  
\usepackage{enumitem}
\usepackage[nolist]{acronym}
\MakePerPage{footnote}
\newcommand{\maz}{\mathcal{Z}}

\newcommand{\man}{\mathcal{N}}

\newcommand{\mad}{\mathcal{D}}

\newcommand{\mai}{\mathcal{I}}

\newcommand{\mak}{\mathcal{K}}

\newcommand{\setR}{\mathbbmss{R}}
\newcommand{\setX}{\mathbbmss{X}}

\newcommand{\setZ}{\mathbbmss{Z}}

\newcommand{\rmp}{\mathrm{p}}

\newcommand{\bx}{{\boldsymbol{x}}}
\newcommand{\hx}{{\hat{x}}}

\newcommand{\bhx}{{\boldsymbol{\hat{x}}}}

\newcommand{\bxx}{\mathbf{x}}
\newcommand{\bvv}{\mathbf{v}}

\newcommand{\bz}{{\boldsymbol{z}}}

\newcommand{\bg}{{\mathbf{g}}}

\newcommand{\bv}{{\boldsymbol{v}}}

\newcommand{\dif}{\mathrm{d}}

\newcommand{\trp}{\mathsf{T}}

\newcommand{\sfr}{\mathsf{r}}
\newcommand{\sfd}{\mathsf{d}}
\newcommand{\sfp}{\mathsf{p}}
\newcommand{\sfD}{\mathsf{D}}

\newcommand{\by}{{\boldsymbol{y}}}

\newcommand{\mA}{\mathbf{A}}
\newcommand{\mI}{\mathbf{I}}
\newcommand{\mone}{\mathbf{1}}
\newcommand{\mJ}{\mathbf{J}}

\newcommand{\mQQ}{\mathbf{Q}}

\newcommand{\mU}{\mathbf{U}}

\newcommand{\mD}{\mathbf{D}}
\newcommand{\mM}{\mathbf{M}}

\newcommand{\rmF}{\mathrm{F}}

\newcommand{\rmg}{\mathrm{g}}
\newcommand{\mT}{\mathbf{T}}
\newcommand{\md}{\mathrm{D}}
\newcommand{\E}{\mathsf{E}}

\newcommand{\abs}[1]{\lvert #1 \rvert}

\newcommand{\set}[1]{\left\lbrace #1 \right\rbrace}
\newcommand{\norm}[1]{\lVert #1 \rVert}

\newtheoremstyle{mystyle}%                % Name
  {}%                                     % Space above
  {}%                                     % Space below
  {}%                                     % Body font
  {}%                                     % Indent amount
  {\bfseries}%                            % Theorem head font
  {:}%                                     % Punctuation after theorem head
  { }%                                    % Space after theorem head, ' ', or \newline
  {}%                                     % Theorem head spec (can be left empty, meaning `normal')

\theoremstyle{mystyle}

\newtheorem*{definition}{Definition}
\newtheorem*{prf}{Sketch of the proof}
\newtheorem{proposition}{Proposition}

\newtheorem*{rs}{RS Ansatz}

\newtheorem*{rsb1}{One-step RSB Ansatz}

\newcounter{bar}

\begin{document}
\title{Replica~Symmetry~Breaking~in~Compressive~Sensing}

\author{\IEEEauthorblockN{Ali Bereyhi\IEEEauthorrefmark{1}, Ralf M\"uller\IEEEauthorrefmark{1}, Hermann Schulz-Baldes\IEEEauthorrefmark{2}}
\IEEEauthorblockA{\IEEEauthorrefmark{1}Institute for Digital Communications (IDC), \IEEEauthorrefmark{2}Department of Mathematics,\\ Friedrich Alexander University (FAU), Erlangen, Germany\\Email: ali.bereyhi@fau.de, ralf.r.mueller@fau.de, schuba@mi.uni-erlangen.de}
\thanks{This work was supported by the German Research Foundation, Deutsche Forschungsgemeinschaft (DFG), under Grant No. MU 3735/2-1.}
}
%\and
%\IEEEauthorblockN{Homer Simpson}
%\IEEEauthorblockA{Twentieth Century Fox\\
%\and
%\IEEEauthorblockN{James Kirk\\ and Montgomery Scott}
%\IEEEauthorblockA{Starfleet Academy\\
%San Francisco, California 96678--2391}}


%Telephone: (800) 555--1212, Fax: (888) 555--1212}
\IEEEoverridecommandlockouts
\maketitle
\begin{abstract}
For noisy compressive sensing systems, the asymptotic distortion with respect to an arbitrary distortion function is determined when a general class of least-square based reconstruction schemes is employed.  The sampling matrix is considered to belong to a large ensemble of random matrices including i.i.d. and projector matrices, and the source vector is assumed to be i.i.d. with a desired distribution. We take a statistical mechanical approach by representing the asymptotic distortion as a macroscopic parameter of a spin glass and employing the replica method for the large-system analysis. In contrast to earlier studies, we evaluate the general replica ansatz which includes the RS ansatz as well as RSB. The generality of the solution enables us to study the impact of symmetry breaking. Our numerical investigations depict that for the reconstruction scheme with the ``zero-norm’’ penalty function, the RS fails to predict the asymptotic distortion for relatively large compression rates; however, the one-step RSB ansatz gives a valid prediction of the performance within a larger regime of compression rates.
\end{abstract}
% \fi
%\IEEEpeerreviewmaketitle
%\vspace{-1mm}
\section{Introduction}
The vector-valued linear system 
\begin{align}
\by=\mA \bx + \bz \label{eq:1}
\end{align}
describes a sampling system in which the source vector $\bx_{n \times 1} \in \setX^n$ with $\setX\subseteq\setR$ is linearly measured by the sampling matrix $\mA_{k \times n}\in \setR^{k \times n}$ and corrupted by zero-mean additive white Gaussian noise $\bz_{k \times 1} \sim \man(\boldsymbol{0}, \lambda_0 \mI)$. The source vector is reconstructed from the observation vector $\by_{k \times 1}$ using the least-square based reconstruction scheme with% function $u(\cdot)$ which reads
\begin{align}
\bg(\by) \coloneqq \arg \min_{\bv\in\setX^n} \ \left[ \frac{1}{2\lambda} \norm{\by-\mA \bv}^2 + u(\bv) \right] \label{eq:4}
\end{align}
for some general penalty function $u(\cdot)$ and tuning factor $\lambda$. The reconstruction scheme in \eqref{eq:4} can be considered as a \ac{map} estimator which postulates the prior distribution to be proportional to $e^{-u(\bx)}$ and the noise variance to be $\lambda$. The optimality of the scheme, therefore, depends on the choice of $u(\cdot)$ and $\lambda$. In compressive sensing, the source vector is sparse meaning that it contains a certain number of zero entries \cite{donoho2006compressed,candes2006robust}. The typical choice for the penalty function in this case is an $\ell_\sfp$-norm. Different choices of $\sfp$ result in various levels of optimality and complexity which always contain a tradeoff in between; the better the scheme performs, the more complex it is. For noisy sampling systems, the performance of the reconstruction scheme is quantified by the average distortion which reads
\begin{align}
\sfD_n = \frac{1}{n} \sum_{j=1}^n \sfd(x_j;\hx_j). \label{eq:ave_dist}
\end{align}
for some general distortion function $\sfd(\cdot;\cdot): \setX \times \setX \mapsto \setR$, and $\bhx_{n\times1}=\bg(\by)$. In the literature, the most trivial choices for $\ell_\sfp$ are the $\ell_2$-norm, $\ell_1$-norm and zero-norm which respectively correspond to the ``linear'', ``LASSO'' \cite{tibshirani1996regression} and ``zero-norm'' reconstruction schemes. The former two choices of $\sfp$ result in convex optimization problems which make them computationally feasible. The latter scheme, however, confronts a non-convex and computationally unfeasible problem. We are interested in studying the asymptotic performance of the general reconstruction scheme given in \eqref{eq:4} when the dimensions grow large. The analysis strategy in this case is to consider a random sampling matrix and determine the average distortion for a given realization of it. In this case, the asymptotic performance is evaluated by taking the expectation over the matrix distribution first, and then, taking the limit $n, k \uparrow\infty$. This task is not trivial for most cases of the function $u(\cdot)$ and the support $\setX$, and therefore, many analytical methods fail. An alternative approach is based on statistical mechanics in which the asymptotics of the sampling system are represented as macroscopic parameters of a spin glass \cite{edwards1975theory}. In this paper, we take the latter approach and invoke the replica method to study the asymptotics of the reconstruction scheme~given~in~\eqref{eq:4}.

\subsection*{Replica Method and its Applications}
The replica method is a nonrigorous but effective method developed in the physics literature to study spin glasses. Although the method lacks rigorous mathematical proof in some particular parts, it has been widely accepted as an analysis tool and utilized to investigate a variety of problems in applied mathematics, information processing, and coding \cite{mezard1986replica,fu1986application,nishimori2001statistical,montanari2000turbo,watkin1993statistical,kabashima1999statistical}. Regarding multiuser communication systems, the method was initially employed by Tanaka to investigate the asymptotic performance of randomly spread CDMA multiuser detectors \cite{tanaka2002statistical}. For communication systems of form \eqref{eq:1} with an \ac{iid} matrix, the authors in \cite{guo2005randomly} considered a class of postulated minimum \ac{mse} estimators and extended the earlier analyses to a larger set of input distributions. The study, moreover, justified the decoupling property of the postulated minimum \ac{mse} which was earlier conjectured in \cite{muller2002channel} and indicates that the pair of input-output symbols are asymptotically converging in distribution to the input-output symbols of an equivalent single-user system. The characteristics of the equivalent system were then determined through the replica analysis. Due to the similarity between the \ac{map} estimation and sampling systems' reconstruction schemes, the replica method has been further used to study compressive sensing \cite{guo2009single,wu2012optimal}. The authors of \cite{rangan2012asymptotic} extended the scope of the decoupling property to a large class of \ac{map} estimators by representing the \ac{map} estimator as the limit of a sequence of minimum \ac{mse} etimators and using the replica results of \cite{guo2005randomly}. The result was then employed to study the asymptotics of $\ell_2$-, $\ell_1$- and zero-nerm based reconstructions in compressive sensing systems. The asymptotic \ac{mse} of regularized least-square reconstruction schemes was, moreover, determined in \cite{vehkapera2014analysis} for a wider range of matrices. In \cite{tulino2013support}, the problem of support recovery was considered where the authors determined the asymptotic input-output information rate and support recovery error for a class of sampling systems. The aforementioned studies were considered under the \ac{rs} assumption which assumes the equivalent spin glass to have some symmetric properties. Although the \ac{rs} assumption has been successful in tracking some solutions, there exist several examples in which it clearly fails. In \cite{zaidel2012vector}, the authors showed that the earlier \ac{rs}-based investigations of vector precoding in \cite{muller2008vector} clearly violates the theoretically rigorous lower bound for some example of lattice precoding. They, therefore, employed Parisi's scheme of \ac{rsb} \cite{parisi1980sequence} to determine a more general ansatz through the replica analysis. The result depicted that the performance prediction via one-step of \ac{rsb} is consistent with the theoretical bounds given in the literature. Inspired by \cite{zaidel2012vector}, the \ac{map} estimator was investigated in \cite{bereyhi2016itw} under \ac{rsb} and it was shown that the \ac{rs} decoupling property reported in \cite{rangan2012asymptotic} holds in a more general form under the \ac{rsb} assumption. The investigations of the least-square error precoding also has shown several examples in which the \ac{rs} assumption results in a theoretically invalid solution, and therefore, the \ac{rsb} ans\"atze were needed for assessing the performance \cite{sedaghat2016lse}. Regarding the compressive sensing systems, the stability analysis of $\ell_\sfp$-norm based reconstruction schemes in \cite{kabashima2009typical} for the noiseless sampling systems has shown that in contrast to the convex cases of $\ell_2$- and $\ell_1$-norm, the \ac{rs} ansatz for the zero-norm based scheme is not locally stable against perturbations that break the symmetry of the replica correlation matrix. The fact which resulted in the conclusion that for this case the \ac{rsb} ans\"atze are required for precise approximation of the asymptotic performance.

\subsection*{Contributions and Organization}
This paper determines the asymptotic distortion of the reconstruction scheme \eqref{eq:4} when it is employed for recovering the source vector from the noisy sampling system \eqref{eq:1} via the replica method. The distortion function, as well as the source distribution, is considered to be general, and the sampling matrix $\mA$ belongs to a wide set of random ensembles. We deviate from the earlier replica analyses of compressive sensing systems by evaluating the general replica ansatz which includes all the possible structures for the replica correlation matrix. The generality of the replica ansatz enables us to determine the \ac{rs} as well as \ac{rsb} ansatz as special cases, and therefore, investigate the impact of symmetry breaking. The analytical results in special cases recover the earlier \ac{rs} based studies of compressive sensing systems, e.g.,  \cite{rangan2012asymptotic,vehkapera2014analysis,tulino2013support,guo2009single}. Moreover, our numerical investigations show that for the zero-norm reconstruction, the \ac{rs} ansatz fails to predict the performance for relatively large compression rates while the \ac{rsb} ans\"atze approximate the performance validly. An introduction to the replica method, is given through the asymptotic analyses in Section \ref{sec:2-2}.
\subsection*{Notation}
We represent scalars, vectors and matrices with non-bold, bold lower case and bold upper case letters, respectively. A $k \times k$ identity matrix is shown by $\mI_k$, and the $k \times k$ matrix with all entries equal to one is denoted by $\mone_k$. $\mA^{\trp}$ indicates the Hermitian of the matrix $\mA$. The set of real and integer numbers are denoted by $\setR$ and $\setZ$, and their corresponding non-negative subsets by superscript $+$. $\norm{\cdot}$ and $\norm{\cdot}_1$ denote the $\ell_2$- and $\ell_1$-norm respectively, and $\norm{\bx}_0$ represents the zero-norm defined as the number of nonzero entries. For a random variable $x$, $\mathrm{p}_x$ represents either the \ac{pmf} or \ac{pdf}, and $\rmF_x$ identifies the \ac{cdf}. Moreover, $\E_x$ identifies mathematical expectation over $x$, and an expectation over all random variables involved in a given expression is denoted by $\mathsf{E}$. For sake of compactness, the set of integers $\set{1, \ldots , n}$ is abbreviated as $[1:n]$ and a zero-mean and unit-variance Gaussian distribution is represented by $\phi(\cdot)$. Gaussian averages are shown as 
\begin{align}
\int ( \cdot ) \ \md z = \int ( \cdot ) \ \frac{e^{-\frac{{z^2}}{2}}}{\sqrt{2 \pi}} \ \dif z.
\end{align}
Whenever needed, we assume the support $\setX$ to be discrete. The results, however, are in full generality and hold also for continuous distributions.
%
%Throughout the paper, we represent vectors with bold lower case letters, scalars with non-bold lower case letters, and matrices with bold upper case letters. The set of real numbers is denoted by $\mathsf{R}$, and $\mA^{\trp}$ indicates the transposed of $\mA$. Furthermore,  $\mI_m$ is the $m\times m$ identity matrix and $\mone_m$ is the matrix with all entries equal to one. For a random variable $x$, we denote the \ac{cdf}, \ac{pmf}, \ac{pdf}, and mathematical expectation over $x$, by $\mathrm{F}_x$, $\mathrm{P}_x$, $\mathrm{f}_x$, $\mathsf{E}_x$, respectively. An expectation over all random variables involved in a given expression is denoted by $\mathsf{E}$. Gaussian averages are abbreviated as 
%\begin{align}
%\int ( \cdot ) \ \md z = \int ( \cdot ) \ \frac{e^{-\frac{{z^2}}{2}}}{\sqrt{2 \pi}} \ \dif z.
%\end{align}
%For sake of simplicity, we restrict to a discrete support $\maX\subset\mathsf{R}$, but with adequate modifications the results and techniques extend to continuous distributions.
\section{Problem Formulation}
\label{sec:2}
Suppose $\by_{k \times 1}$ is given by a sampling system as in \eqref{eq:1} where
\begin{enumerate}[label=(\alph*)]
\item $\bx_{n \times 1}$ is an \ac{iid} random vector with each entry~being~distributed with $\rmp_x$ over $\setX\subseteq\setR$.
\item $\mA$ is a ${k \times n}$ random matrix over $\setR^{k \times n}$, such that its Gramian $\mJ\coloneqq \mA^{\trp} \mA$ has the eigendecomposition 
\begin{align}
\mJ= \mU \mD \mU^{\trp} \label{eq:2}
\end{align}
with $\mU$ being an orthogonal Haar distributed matrix and $\mD$ being a diagonal matrix of which the empirical \ac{cdf} of eigenvalues (density of states) converges as $n\uparrow\infty$ to a deterministic \ac{cdf} $\mathrm{F}_{\mJ}$.
\item $\bz_{k \times 1}$ is a real \ac{iid} zero-mean Gaussian random vector with variance $\lambda_0$, i.e., $\bz \sim \man(\boldsymbol{0},\lambda_0 \mI)$.
\item The number of observations $k$ is a deterministic function of the system dimension $n$ such that
\begin{align}
\lim_{n \uparrow \infty} \frac{k(n)}{n}=\frac 1 \sfr < \infty. \label{eq:3}
\end{align}
\item $\bx$, $\mA$ and $\bz$ are independent.
\end{enumerate}
We reconstruct the source vector $\bx$ from $\by$ as $\bhx=\bg(\bx)$ with $\bg(\cdot)$ being defined in \eqref{eq:4} and satisfies the following constraints.
\begin{enumerate}[label=(\alph*)]
\item The penalty function $u(\cdot)$ decouples meaning that
\begin{align}
u(\bv)=\sum_{i=1}^k u(v_i).
\end{align}
\item For a given vector $\by$, the objective function in \eqref{eq:4} has a unique minimizer over the support $\setX^n$.
\end{enumerate}
%The observation vector $\by$ is the input to the \ac{map} estimator $\bhx(\by)$ which might be mismatched in the sense that the priori distribution is not proportional to $e^{-u(\bx)}$ and/or $\lambda$ in \eqref{eq:4} differs from $\lambda_0$.
%Given the system matrix $\mA$, the source vector is reconstructed from the observation vector $\mA$ as in \eqref{eq:4} for some penalty function $u(\cdot)$ which decouples, i.e.,
%\begin{align}
%u(\bv)=\sum_{i=1}^k u(v_i).
%\end{align}
%Thus, for a given $n$, the average distortion $\mathsf{D}_n$, regarding some distortion function $\sfd(\cdot;\cdot)$ defined in \eqref{eq:dist_fun}, is determined as in \eqref{eq:ave_dist}.
%\begin{align}
%\mathsf{D}_n=\frac{1}{n} \ \E \mathsf{d} (\bhx(\by);\bx). \label{eq:5}
%\end{align}
%The asymptotic distortion, normalized by the system's dimension, is now determined as
%\begin{align}
%\mathsf{D}=\lim_{n \uparrow \infty} \mathsf{D}_n
%\end{align}
%Here, we assume the distortion measure $\mathsf{d}(\cdot;\cdot)$ to be a non-negative function and have the decoupling property, i.e., $\mathsf{d}(\bhx;\bx)= \sum_{i=1}^n \mathsf{d}(\hx_i;x_i)$. The results, however, can easily be extended to larger classes of distortion measures.
Our goal is to determine the asymptotic average distortion $\sfD$ for this setup which is defined as the large limit of the expected average distortion defined in \eqref{eq:ave_dist}, i.e., $\sfD=\lim_{n\uparrow\infty} \E\hspace*{.7mm}\sfD_n$. To do so, we represent $\sfD$ as the macroscopic parameter of a spin glass and invoke the replica method to determine that.
\section{Statistical Mechanical Approach}
\label{sec:2-2}
Consider a spin glass with the Hamiltonian
\begin{align}
\mathcal{E}(\bv|\by, \mA)= \frac{1}{2\lambda} \norm{\by-\mA \bv}^2 + u(\bv) \label{eq:6}
\end{align}
for given $\by$ and $\mA$. At the inverse temperature $\beta$, the microstate $\bv$ is distributed with
\begin{align}
\rmp^{\beta}(\bv)= \frac{e^{-\beta \mathcal{E}(\bv|\by, \mA)}}{\sum_{\bv} e^{-\beta\mathcal{E}(\bv|\by, \mA)}}. \label{eq:9}
\end{align}
By using a standard large deviation argument and defining 
\begin{align}
\mathsf{d} (\bv;\bx)=\sum_{j=1}^n \mathsf{d} (v_j;x_j),
\end{align}
it is shown that the asymptotic distortion reads
\begin{align}
\mathsf{D}=\lim_{\beta \uparrow \infty} \lim_{n \uparrow \infty} \frac{1}{n} \hspace{1mm} \E\set{\mathsf{E}^\beta_{\bv} \hspace{1mm} \mathsf{d} (\bv;\bx)} \label{eq:10}
\end{align}
where $\mathsf{E}^\beta_{\bv}$ takes expectation over the vector $\bv$ \ac{wrt} $\rmp^{\beta}$. \eqref{eq:10} describes $\sfD$ as a macroscopic parameter of the spin glass specified by the Hamiltonian in \eqref{eq:6}. At this point, one utilizes a common trick in statistical mechanics which defines the ``modified partition function'' as
\begin{align}
\maz(\beta, h|\by, \mA)= \sum_{\bv} e^{-\beta \mathcal{E}(\bv|\by, \mA)+h \mathsf{d}(\bv,\bx)}, \label{eq:11}
\end{align}
and determines the macroscopic parameter as
%\begin{align}
%\mathsf{E}^\beta_{\bv} \hspace{1mm} \mathsf{d} (\bv;\bx) = \frac{\partial}{\partial h}  \log \maz(\beta, h|\by, \mA)|_{h=0} \label{eq:12}
%\end{align}and therefore, \eqref{eq:10} reduces to the evaluation of
\begin{align}
\mathsf{D}=\lim_{\beta \uparrow \infty} \lim_{n \uparrow \infty} \frac{1}{n} \frac{\partial}{\partial h} \ \E \log \maz(\beta, h|\by, \mA)|_{h=0}. \label{eq:13}
\end{align}
\eqref{eq:13} raises the nontrivial problem of determining a logarithmic expectation. Here, one may take a step further and employ the Riesz equality which for a given random variable $t$ states 
\begin{align}
\E \log t = \lim_{m \downarrow 0} \frac{1}{m} \log \E t^m,
\end{align}
and write \eqref{eq:13} as
\begin{align}
\mathsf{D}= \lim_{\beta \uparrow \infty} \lim_{n \uparrow \infty}\lim_{h \downarrow 0 }\lim_{m \downarrow 0}\frac{1}{n} \frac{\partial}{\partial h} \frac{1}{m} \log \E [\maz(\beta, h|\by, \mA)]^m.  \label{eq:15}
\end{align}
Determining the \ac{rhs} of \eqref{eq:15} faces two main difficulties. In fact, one needs to evaluate the moment for any real value of $m$ (or at least in the right neighborhood of $0$), and also take the limits in the order stated. This is where the replica method plays its role. It considers the expression under the logarithm in the \ac{rhs} of \eqref{eq:15} as a function in terms of $m$, namely $f(m)$, and conjectures that 
\begin{enumerate}
\item the analytic continuation of $f(\cdot)$ from $\setZ^+$ onto $\setR^+$ equals to $f(m)$ which intuitively states that the final expression of $f(m)$ determined for $m\in\setZ^+$ is same as $f(m)$ for real values of $m$, and
\item the limits with respect to $m$ and $n$ exchange.
\end{enumerate}
The conjecture is known as ``replica continuity'' and is where the replica method lacks rigorousness. By the replica continuity conjecture, $f(m)$ reads
%\begin{assumption}[Replica Continuity] \label{assumption:1}
%For a thermodynamic system with Hamiltonian as in \eqref{eq:6}, replica continuity holds.
%\end{assumption}
%&= \E [\sum_{\bv} e^{-\beta \mathcal{E}(\bv|\by, \mA)+h \mathsf{d}(\bv,\bx)}]^m \nonumber \\
%For integer $m$, the expectation on the \ac{rhs} of \eqref{eq:15} is evaluated as
\begin{align}
f(m)&\coloneqq\E [\maz(\beta, h|\by, \mA)]^m \nonumber \\
&= \E \prod_{a=1}^m \sum_{\bv_a} e^{-\beta \mathcal{E}(\bv_a|\by, \mA)+h \mathsf{d}(\bv_a,\bx)}. \label{eq:16}
\end{align}
%Here, the \ac{rhs} of \eqref{eq:16} is a product of $m$ copies of the partition function, or equivalently, can be interpreted as the partition function of a new thermodynamic system being constructed by collecting $m$ copies of the original system, independently. In fact, this is where the terminology ``replica'' comes from. 
%Starting from \eqref{eq:16}, by several lines of derivations Proposition \ref{proposition:1} is concluded.

\section{Main Results}
\label{sec:3}
Proposition \ref{proposition:1} gives the general replica ansatz which only relies on the replica continuity conjecture. Before stating the proposition, let us define the $\mathrm{R}$-transform.
\begin{definition}
Considering a random variable $t\sim\rmp_t$, the corresponding Stieltjes transform over the upper half complex plane is defined as $\mathrm{G}_t(s)= \E (t-s)^{-1}$. Denoting the inverse \ac{wrt} composition by $\mathrm{G}_t^{-1}(\cdot)$, the $\mathrm{R}$-transform is given by
\begin{align}
\mathrm{R}_t(\omega)= \mathrm{G}_t^{-1}(\omega) - \omega^{-1} \label{eq:19}
\end{align}such that $\lim_{\omega \downarrow 0} \mathrm{R}_{t}(\omega) = \E t$. The definition can be also extended to matrix arguments. Assuming a matrix $\mM_{n \times n}$ to have the eigendecomposition $\mM=\mU \ \mathrm{diag}[\lambda_1, \ldots, \lambda_n] \ \mU^{\trp}$, $\mathrm{R}_t(\mM)$ is then defined as $\mathrm{R}_t(\mM)=\mU \ \mathrm{diag}[\mathrm{R}_t(\lambda_1), \ldots, \mathrm{R}_t(\lambda_n)] \ \mU^{\trp}$.
\end{definition}

\begin{proposition}[General Replica Ansatz]
\label{proposition:1}
Let the linear system \eqref{eq:1} fulfill the constraints of Section \ref{sec:2}. For non-negative integer $m$, define the function
\begin{align}
\mad(\beta,m) &=  \frac{1}{m} \ \E \mathsf{d} (\bvv; \bxx) \label{eq:20}
\end{align}
where $\bxx_{m \times 1}$ is a vector with all elements equal to the random variable $x\sim\rmp_x$, and $\bvv_{m \times 1}\in \setX^m$ is a random vector with conditional distribution
\begin{align}
\rmp^\beta_{\bvv|\bxx}(\bvv|\bxx)=\frac{e^{-\beta (\bxx-\bvv)^{\trp} \mT \mathrm{R}_{\mJ}(-2 \beta \mT \mQQ) (\bxx-\bvv) - \beta u(\bvv) }}{\sum_{\bvv} e^{-\beta (\bxx-\bvv)^{\trp} \mT \mathrm{R}_{\mJ}(-2\beta \mT \mQQ) (\bxx-\bvv) - \beta u(\bvv) }}. \label{eq:21}
\end{align}
In \eqref{eq:21}, $\mathrm{R}_{\mJ} ( \cdot)$ is the $\mathrm{R}$-transform corresponding to $\mathrm{F}_{\mJ}$,
\begin{align}
\mT_{m \times m} = \frac{1}{2 \lambda} \mI_m - \beta \frac{\lambda_0}{2 \lambda^2} \mone_m,
\end{align}
and $\mQQ$ is the so-called replica correlation matrix which satisfies the fixed point equation
\begin{align}
\mQQ= \E\hspace*{1mm} \mathsf{E}_{{\rmp}^\beta_{\bvv|\bxx}} (\bxx - \bvv)(\bxx - \bvv)^{\trp} \label{eq:22}
\end{align}
where $\mathsf{E}_{{\rmp}^\beta_{\bvv|\bxx}}$ takes expectation over $\bvv$ \ac{wrt} ${\rmp}^\beta_{\bvv|\bxx}$.
Then, under the replica continuity conjecture, the asymptotic average distortion is given by
\begin{align}
\mathsf{D}= \lim_{\beta\uparrow\infty}\lim_{m\downarrow 0} \mad(\beta,m). \label{eq:pro1}
\end{align}
\end{proposition}
%\textbf{Sketch of Proof:} The proof has been omitted due to the lack of place; however, we try to give a brief sketch of the proof, here. Starting from the term in \eqref{eq:16}, we first take the expectation \ac{wrt} the noise vector a
%Due to the lack of place, the proof of the proposition, as well as the corollaries which are stated next, is omitted and given in extended versions of the manuscript. However, let us just briefly mention the strategy.
%
%\vspace{.2cm}
\begin{prf}
Starting from \eqref{eq:16} and after evaluating the expectations \ac{wrt} $\bz$ and $\mA$, the \ac{lhs} of \eqref{eq:16} is expressed in terms of the replica correlation matrix $\mQQ_{m\times m}$ whose entries are defined as
\begin{align}
[\mQQ]_{ab} = \frac{1}{n} (\bx-\bv_a)^\trp (\bx-\bv_b). \label{eq:17}
\end{align}
Taking the limits $n\uparrow\infty$ and $h\downarrow 0$, one is lead to use the saddle point method. Finally, using the law of large numbers, the equations in Proposition \ref{proposition:1} are obtained. The detailed derivations are given in \cite{bereyhi2016statistical}.\hfill $\blacksquare$
\end{prf}
Solving the fixed point equation \eqref{eq:22} is notoriously difficult and possibly not of use, because it may depend in a non-analytic way on $m$. To address both issues, one restricts the search of the fixed point solutions to a small parameterized set of correlation matrices. In the sequel, we treat some of the well-known sets.

\subsection{\ac{rs} Ansatz} 
\ac{rs} assumes that the valid solution of the fixed-point equation \eqref{eq:22} is invariant under all permutations of the $m$ replica indices, namely $\mathbf{\Pi}^{-1} \mQQ\mathbf{\Pi}=\mQQ$ for all permutation matrices $\mathbf{\Pi}$ taken from the symmetric group on $[1:m]$. This implies that $\mQQ$ is of the form
\begin{align}
\mQQ = q \mone_m + \frac{\chi}{\beta} \mI_m \label{eq:23}
\end{align}for some non-negative real $q$ and $\chi$. Indeed, this leads to an analytic expression for \eqref{eq:20}, and therefore, the limit in \eqref{eq:pro1} is determined which concludes the following ansatz.

\begin{rs}
\label{corollary:1}
Define $\xi \coloneqq \lambda \ [\mathrm{R}_\mJ (-\dfrac{\chi}{\lambda})]^{-1}$ and $f$ as
\begin{align}
&f \coloneqq \frac{1}{\mathrm{R}_\mJ(-\dfrac{\chi}{\lambda})} \ \sqrt{\frac{\partial}{\partial \chi}[(\lambda_0 \chi - \lambda q) \mathrm{R}_\mJ (-\frac{\chi}{\lambda})]}
\end{align}
for some $\chi$ and $q$. Moreover, let
\begin{align}
\rmg\left(x,z\right)\coloneqq \arg \min_{v} \left[ \frac{1}{2 \xi} \abs{x+ fz -v}^2 + u(v) \right]. \label{eq:26}
\end{align}
Then, the \ac{rs} prediction of $\sfD$ is given by
\begin{align}
\mathsf{D}= \mathsf{E} \int \mathsf{d}\left(\rmg\left(x,z\right);x\right) \ \md z,
\end{align}
for $x\sim\rmp_x$, and $\chi$ and $q$ satisfying
\begin{align}
\chi &=  \frac{\xi}{f} \ \mathsf{E}_x \int \left(\rmg\left(x,z\right)-x\right) z \ \md z, \nonumber \\
q &= \mathsf{E}_x \int \left(\rmg\left(x,z\right)-x\right)^2 \ \md z. \label{eq:29}
\end{align}
\end{rs}
The postulated symmetry of the ansatz, assumed in \ac{rs}, does not necessarily hold, and therefore, the \ac{rs} ansatz may fail to give a valid prediction of $\sfD$.
%Corollary \ref{corollary:1} may fail to give a valid prediction of the asymptotic distortion in some cases. On the other hand, Proposition \ref{proposition:1} allows us to impose some other structures on $\mQQ$ which may avoid the possible error of the \ac{rs} ansatz. In \cite{parisi1980sequence}, Parisi introduced the \ac{rsb} scheme for defining a larger space of correlation matrices recursively. Here, we employ the \ac{rsb} scheme to consider a broader set of correlation matrices and determine the asymptotic distortion for these structures.
%\vspace{-3mm}

\subsection{\ac{rsb} Ans\"atze}
The \ac{rsb} structures are constructed via Parisi's iterative scheme introduced in \cite{parisi1980sequence}. The scheme takes the \ac{rs} correlation matrix as the starting point, and then recursively constructs new structures. After $b$ steps of recursion, which are referred to as breaking steps, the correlation matrix reads
\begin{align}
\mQQ= q \mone_m + \sum_{\kappa=1}^b p_\kappa \mI_{\frac{m \beta}{\mu_\kappa}} \otimes \mone_{\frac{\mu_\kappa}{\beta}} + \frac{\chi}{\beta} \mI_m \label{eq:40}
\end{align}
%
%\begin{align}
%\mQQ= q \mone_m + p \mI_{\frac{m \beta}{\mu}} \otimes \mone_{\frac{\mu}{\beta}} + \frac{\chi}{\beta} \mI_m \label{eq:30}
%\end{align}
for some non-negative $\chi$ and $q$ and sequences $\set{\mu_\kappa}$, and $\set{p_\kappa}$. Similar to the \ac{rs} case, \eqref{eq:40} leads to an analytic~expression~for $\mad(\beta,m)$ which lets us determine $\sfD$ via Proposition \ref{proposition:1}. For sake of compactness, we first state the result for $b=1$, and then illustrate the generalized ansatz with $b$ breaking steps.
\begin{rsb1}
\normalfont
\label{corollary:2}
Define $\xi \coloneqq \lambda \ [\mathrm{R}_\mJ (-\dfrac{\chi}{\lambda})]^{-1}$ and
\begin{subequations}
\begin{align}
f &\coloneqq \frac{1}{\mathrm{R}_\mJ(-\dfrac{\chi}{\lambda})} \sqrt{\frac{\partial}{\partial \varrho} [\lambda_0 \varrho + \lambda p - \lambda q ] \mathrm{R}_{\mJ}(- \frac{\varrho}{\lambda})} \label{eq:31} \\
w &\coloneqq \frac{1}{\mathrm{R}_\mJ(-\dfrac{\chi}{\lambda})} \sqrt{\dfrac{\lambda}{\mu} \hspace{.7mm} \mathrm{R}_{\mJ}(- \frac{\chi}{\lambda}) - \mathrm{R}_{\mJ}(- \frac{\varrho}{\lambda})]} \label{eq:32}
\end{align}
\end{subequations}
for some real $\chi$, $q$, $\mu$ and $p$, and $\varrho\coloneqq \chi+\mu p$; moreover let
\begin{align}
\hspace*{-1.8mm}\mak(v,x,z,y)\hspace*{-.7mm} =\hspace*{-.7mm} \dfrac{1}{2\xi} \left[ \left(x\hspace*{-.7mm}-\hspace*{-.7mm}v\right)^2\hspace*{-.7mm}+\hspace*{-.7mm}2\left( x\hspace*{-.7mm}-\hspace*{-.7mm}v\right)\left( f z \hspace*{-.7mm}+ \hspace*{-.7mm} w y \right)\right]\hspace*{-.7mm}+\hspace*{-.7mm} u(v) \label{eq:34}
\end{align}
and define the functions
\begin{subequations}
\begin{align}
\mathcal{L}\left(x,z,y\right)&= \min_{v} \mak(v,x,z,y) \\
\rmg\left(x,z,y\right)&= \arg \min_{v} \mak(v,x,z,y)\label{eq:gg}
\end{align}
\end{subequations}
Then, the asymptotic average distortion is given by
\begin{align}
\mathsf{D}= \mathsf{E} \int \mathsf{d}\left(\rmg\left(x,z,y\right);x\right) \mai\left(x,z,y\right) \ \md y \md z \nonumber
\end{align}
where $x\sim\rmp_x$ and $\mai\left(x,z,y\right)$ reads
\begin{align}
\mai\left(x,z,y\right) &= \frac{e^{-\mu \mathcal{L}\left(x,z,y\right)}}{\int e^{-\mu \mathcal{L}\left(x,z,y\right)} \md y}, \label{eq:36}
\end{align}
for $\chi$, $p$ and $q$ which satisfy
\begin{subequations}
\begin{align}
\varrho &= \frac{\xi}{f} \hspace*{.7mm} \mathsf{E} \int \left(\rmg\left(x,z,y\right)-x\right) z \hspace*{.9mm}\mai\left(x,z,y\right) \md y \md z, \\
\chi + \mu q &= \frac{\xi}{w} \hspace*{.7mm}\mathsf{E} \int \left(\rmg\left(x,z,y\right)-x\right) y \hspace*{.7mm} \mai\left(x,z,y\right) \md y \md z, 
\end{align}
\begin{align}
q &= \mathsf{E} \int \left(\rmg\left(x,z,y\right)-x\right)^2 \hspace*{.9mm} \mai\left(x,z,y\right) \md y \md z \label{eq:38}
\end{align}
\end{subequations}
and $\mu$ which is a solution to
\begin{align}
\frac{\mu p}{2 \xi} = &\frac{1}{2\lambda} \int_{\chi}^{\varrho} \mathrm{R}_{\mJ}(-\frac{\omega} {\lambda}) \hspace*{.5mm} \dif \omega +\frac{\mu^2 w^2}{2 \xi^2}(p-q) \nonumber \\
&\hspace*{2mm}+\mathsf{E} \int \mai\left(x,z,y\right) \log \mai\left(x,z,y\right) \md y \md z. \label{eq:39}
\end{align}
\end{rsb1}
The one-step \ac{rsb} ansatz is further extended to more steps of breaking. In that case the asymptotic distortion is given by
\begin{align}
\hspace*{-1.5mm}\mathsf{D}\hspace*{-.7mm}=\hspace*{-.7mm}\mathsf{E} \int \mathsf{d}(\tilde{\rmg}(x,z,\set{y_\kappa}_{1}^b);x ) \hspace*{.4mm}\tilde{\mai}(x,z,\set{y_\kappa}_{1}^b) \prod_{\kappa=1}^b \md y_\kappa \md z 
\end{align}
where
\begin{align}
\tilde{\mai}(x,z,\set{y_\kappa}_{1}^b) = \prod_{\kappa=1}^b \left[{\Lambda}_\kappa (x,z, \{ y_\varsigma \}_{\kappa}^b)\right]^{\mu_\kappa}
\end{align}
for a sequence of functions $\set{{\Lambda}_\kappa(\cdot,\cdot,\set{\cdot})}_{\kappa=1}^b$, and $\tilde{\rmg}(\cdot,\cdot,\set{\cdot}_{1}^b)$ being defined as in \eqref{eq:gg} by replacing $wy$ in \eqref{eq:34} by $\sum_{\kappa=1}^b w_\kappa y_\kappa$  for some $\set{w_\kappa}_{\kappa=1}^b$. The explicit expression of the functions and scalar factors are omitted for sake of compactness, and are given in the extended version of the manuscript \cite{bereyhi2016statistical}.

\section{Least-Square Reconstruction Schemes}
\label{sec:4}
In compressive sensing, the source vector is supposed to be sparse which means that a certain fraction of entries are zero. 
%
%As an application of our results, we study the asymptotic properties of a compressive sensing system. For this system, $\bx$ represents the source vector which is sampled by the sensing matrix $\mA$, and corrupted by the \ac{iid} Gaussian noise vector $\bz$. The noisy observation vector $\by$ is then given into a decompressor with decompression rule as in \eqref{eq:4} to recover the source vector. Here, $r$, as defined in \eqref{eq:3}, denotes the inverse compression rate which takes large values when the source vector is sparse enough. In fact, the sparsity of the source vector forces a new set of constraints to our problem, and allows for proper reconstruction in an undersampling regime.
%
To model the sparsity of the source, we set $\rmF_x$ to be
\begin{align}
\mathrm{F}_x(x)= (1-s) \mone\{x \geq 0 \} + s \breve{\mathrm{F}}_x(x). \label{eq:Fx}
\end{align}
with $\mone\set{\cdot}$ being the indicator function, for some \ac{cdf} $\breve{\mathrm{F}}_x(x)$ and $0\leq s \leq 1$. By the law of large numbers, $\bx$ in the asymptotic regime has $(1-s) n$~zeros~and $s n$ non-zero entries which are distributed \ac{wrt} \ac{cdf} $\breve{\mathrm{F}}_x$. From the reconstruction point of view, several schemes can be considered by setting different forms of $u(\cdot)$ in \eqref{eq:4}. In this section, we consider the least-square based reconstruction schemes in compressive sensing and investigate the asymptotic performance. Throughout the investigations, we assume%that
\begin{itemize}
\item $\bx$ is an \ac{iid} zero-mean and unit-variance ``sparse Gaussian'' vector meaning that $\breve{\rmF}_x(x)$ in \eqref{eq:Fx} is a zero-mean and unit-variance Gaussian \ac{cdf}.
\item the distortion function $\sfd(\cdot;\cdot)$ is of the form
\begin{align}
\sfd(\hx;x)=(\hx-x)^2 \label{eq:distortion_fun}
\end{align}
which determines the asymptotic average \ac{mse}.
\item the sampling matrix is either an ``\ac{iid} random'' or a ``random projector'' matrix. In the former case, the entries of $\mA_{k\times n}$ are generated \ac{iid} with zero-mean and variance $k^{-1}$. The asymptotic empirical eigenvalue \ac{cdf} of the Gramian $\mJ$, in this case, follows the Marcenko-Pastur law, and therefore, the $\mathrm{R}$-transform is given by
\begin{align}
\mathrm{R}_{\mJ} (\omega) = \frac{1}{1-\sfr \omega}. \label{eq:cs-21}
\end{align}
The latter case describes a sampling matrix in which the rows are orthogonal.  In this case, the $\mathrm{R}$-transform reads
\begin{align}
\mathrm{R}_{\mJ} (\omega) = \dfrac{\sf \omega - 1 + \sqrt{(\sfr \omega -1)^2+4 \omega}}{2 \omega}. \label{cs:24}
\end{align}
\end{itemize}
To study the asymptotics of least-square based reconstruction schemes, we need to set the penalty function to be one of $\ell_2$-, $\ell_1$-, or zero-norm functions.
\begin{enumerate}[label=(\alph*)]
\item By setting the penalty function to be $u(v)=v^2/2$, the linear least-square scheme is recovered.
\item For $u(v)=\abs{v}$, the reconstruction scheme in \eqref{eq:4} reduces to the LASSO \cite{tibshirani1996regression} or basic pursuit denoising \cite{chen2001atomic} scheme.
\item By considering $u(v)=\mone\set{v\neq 0}$ the zero-norm reconstruction scheme is obtained.
\end{enumerate}

\subsection{Numerical Results}
\begin{figure}[t]
\hspace*{-1.3cm}  
%\centering
\resizebox{1.25\linewidth}{!}{
\pstool[width=.35\linewidth]{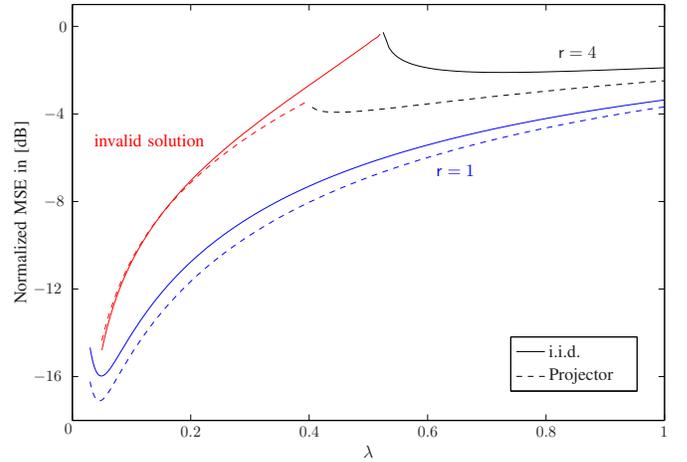}{
\psfrag{D}[c][c][0.2]{\ac{mse} in [dB]}
\psfrag{alpha}[c][c][0.2]{$\lambda$}
\psfrag{D}[c][c][0.2]{Normalized MSE in [dB]}
\psfrag{DDDRRRTTT}[l][l][0.2]{\ac{iid}}
\psfrag{PPPRRRTTT}[l][l][0.2]{Projector}
\psfrag{r=4}[l][l][0.2]{$\sfr=4$}
\psfrag{r=1}[l][l][0.2]{\textcolor{blue}{$\sfr=1$}}
\psfrag{invalidsolution}[l][l][0.2]{\textcolor{red}{invalid solution}}
%y-axis
\psfrag{-5}[r][c][0.18]{$-5$}
\psfrag{-10}[r][c][0.18]{$-10$}
\psfrag{-4}[r][c][0.18]{$-4$}
\psfrag{-8}[r][c][0.18]{$-8$}
\psfrag{-12}[r][c][0.18]{$-12$}
\psfrag{-16}[r][c][0.18]{$-16$}
\psfrag{0}[r][c][0.18]{$0$}
%
%%x-axis
\psfrag{1}[c][b][0.18]{$1$}
\psfrag{0.2}[c][b][0.18]{$0.2$}
\psfrag{0.4}[c][b][0.18]{$0.4$}
\psfrag{0.6}[c][b][0.18]{$0.6$}
\psfrag{0.8}[c][b][0.18]{$0.8$}
\psfrag{2}[c][b][0.18]{$2$}
\psfrag{2.2}[c][b][0.18]{$2.2$}
\psfrag{2.4}[c][b][0.18]{$2.4$}
\psfrag{2.6}[c][b][0.18]{$2.6$}
\psfrag{2.8}[c][b][0.18]{$2.8$}
\psfrag{3}[c][b][0.18]{$3$}
}}
\caption{\ac{rs}-prediction of normalized \ac{mse} vs. the compression rate for the  zero-norm reconstruction scheme. As the compression rate grows, the \ac{rs} fixed-point equation gives invalid predictions. The sparsity factor is considered $s=0.1$ and the noise variance is set $\lambda_0=0.01$.}
\label{fig:2}
\end{figure}
Fig. \ref{fig:2} shows the RS prediction of the normalized \ac{mse} in terms of the tuning factor $\lambda$ for the zero-norm reconstruction. The sparsity factor is considered to be $s=0.1$, and the noise variance is set to be $\lambda_0=0.01$. The curves have been sketched for both the \ac{iid} random and projector sampling matrices at the compression rates $\sfr=1$ and $\sfr=4$. As the figure illustrates, \ac{rs} fails to predict the normalized \ac{mse} at small values of $\lambda$ for large compression rates. In fact, as the compression rate grows, the normalized \ac{mse} drops unexpectedly down for an interval of $\lambda$. This is due to the fact that the \ac{rs} fixed point equations have either an invalid solution or no solution in this interval. In other words, the replica ansatz, for this regime of system parameters, does not exhibit symmetry, and therefore, the \ac{rs} postulated structure for the replica correlation matrix does not lead to the true saddle-point. The result was earlier reported for the noiseless case in \cite{kabashima2009typical} where the authors showed that under a set of constraints the \ac{rs} prediction is not valid for the zero-norm reconstruction.

\begin{figure}[t]
\hspace*{-1.3cm}  
%\centering
\resizebox{1.25\linewidth}{!}{
\pstool[width=.35\linewidth]{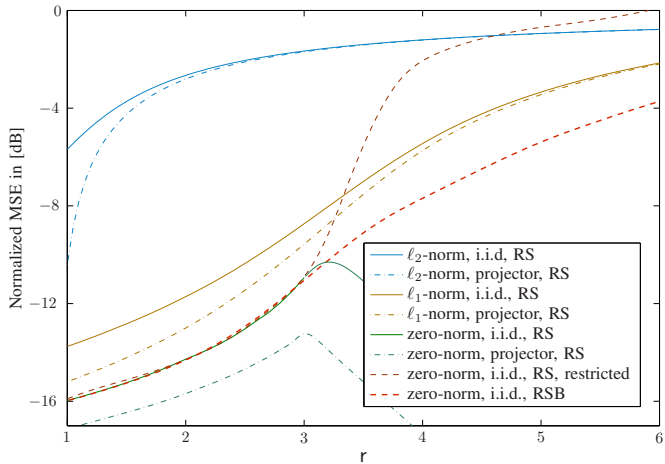}{
\psfrag{D}[c][c][0.2]{Normalized \ac{mse} in [dB]}
\psfrag{alpha}[c][c][0.25]{$\sfr$}
\psfrag{NORM0-XXX-DDDAAABBBAAA}[l][l][0.2]{zero-norm, i.i.d., RS }
\psfrag{NORM1-XXX-DDDAAABBBAAA}[l][l][0.2]{$\ell_1$-norm, i.i.d., RS }
\psfrag{NORM2-XXX-DDDAAABBBAAA}[l][l][0.2]{$\ell_2$-norm, i.i.d, RS }
\psfrag{NORM0-XXX-OOOAAABBBAAA}[l][l][0.2]{zero-norm, projector, RS }
\psfrag{NORM1-XXX-OOOAAABBBAAA}[l][l][0.2]{$\ell_1$-norm, projector, RS }
\psfrag{NORM2-XXX-OOOAAABBBAAA}[l][l][0.2]{$\ell_2$-norm, projector, RS }
\psfrag{NORM0-XXB-DDDAAABBBAAA}[l][l][0.2]{zero-norm, i.i.d., RSB }
\psfrag{NORM0-XXB-DDDAAABBBRRR}[l][l][0.2]{zero-norm, i.i.d., RS, restricted }
%y-axis
\psfrag{-5}[r][c][0.18]{$-5$}
\psfrag{-10}[r][c][0.18]{$-10$}
\psfrag{-4}[r][c][0.18]{$-4$}
\psfrag{-8}[r][c][0.18]{$-8$}
\psfrag{-12}[r][c][0.18]{$-12$}
\psfrag{-16}[r][c][0.18]{$-16$}
\psfrag{0}[r][c][0.18]{$0$}
%
%%x-axis
\psfrag{1}[c][b][0.18]{$1$}
\psfrag{5}[c][b][0.18]{$5$}
\psfrag{4}[c][b][0.18]{$4$}
\psfrag{6}[c][b][0.18]{$6$}
\psfrag{1.8}[c][b][0.18]{$1.8$}
\psfrag{2}[c][b][0.18]{$2$}
\psfrag{2.2}[c][b][0.18]{$2.2$}
\psfrag{2.4}[c][b][0.18]{$2.4$}
\psfrag{2.6}[c][b][0.18]{$2.6$}
\psfrag{2.8}[c][b][0.18]{$2.8$}
\psfrag{3}[c][b][0.18]{$3$}
}}
\caption{\ac{rs} and one-step \ac{rsb} prediction of the normalized \ac{mse} vs. $\sfr$ for $s=0.1$ and $\lambda_0=0.01$. At higher compression rates, the \ac{rs} predicted \ac{mse} unexpectedly drops down. The one-step \ac{rsb}, however, tracks the curve for $\ell_1$-norm reconstruction within a gap.}
\label{fig:1}
\end{figure}
To investigate the impact of \ac{rsb}, the \ac{rs} as well as one-step \ac{rsb} prediction of the normalized \ac{mse} has been plotted in terms of the compression rate in Fig. \ref{fig:1} for the zero-norm reconstruction when $s=0.1$ and $\lambda_0=0.01$. The normalized \ac{mse} has been numerically minimized over $\lambda$. As a benchmark, the \ac{rs} predicted curves for the $\ell_2$- and $\ell_1$-norm schemes have also been sketched. As the figure shows, the \ac{rs} predicted \ac{mse} starts to decrease in higher compression rate regimes and even violates theoretical lower bounds. The one-step \ac{rsb} ansatz, however, is consistent with theoretical bounds, and tracks the curve for the $\ell_1$-norm scheme within a certain gap. For sake of comparison, we have also plotted the ``restricted \ac{rs} prediction''. For this curve, we have minimized the \ac{rs}-predicted normalized \ac{mse} within the interval of $\lambda$ in which the \ac{rs} ansatz is stable. As it is observed, the curve deviates the one-step \ac{rsb} curve, as $\sfr$ grows large. It also violates the $\ell_1$- and $\ell_2$-norm curves at higher compression rates; the fact which indicates that the optimal tuning factor lies within the interval of $\lambda$ with an unstable \ac{rs} ansatz.

\begin{figure}[t]
\hspace*{-1.3cm}  
%\centering
\resizebox{1.25\linewidth}{!}{
\pstool[width=.35\linewidth]{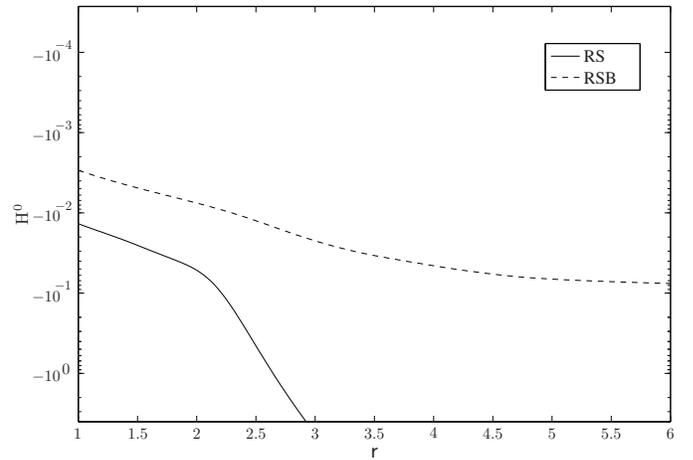}{
\psfrag{D}[c][c][0.2]{$\mathrm{H}^0$}
\psfrag{alpha}[c][c][0.25]{$\sfr$}
\psfrag{Ent-0RS}[l][l][0.2]{RS}
\psfrag{Ent-RSB}[l][l][0.2]{RSB}
\psfrag{NORM2-XXX-DDDAAABBB}[l][l][0.2]{$\ell_2$-norm, i.i.d, RS }
\psfrag{NORM0-XXX-OOOAAABBB}[l][l][0.2]{zero-norm, projector, RS }
\psfrag{NORM1-XXX-OOOAAABBB}[l][l][0.2]{$\ell_1$-norm, projector, RS }
\psfrag{NORM2-XXX-OOOAAABBB}[l][l][0.2]{$\ell_2$-norm, projector, RS }
\psfrag{NORM0-XXB-DDDAAABBB}[l][l][0.2]{zero-norm, i.i.d., RSB }
%y-axis
\psfrag{-0.02}[r][c][0.18]{$-0.02$}
\psfrag{-0.04}[r][c][0.18]{$-0.04$}
\psfrag{-0.06}[r][c][0.18]{$-0.06$}
\psfrag{-0.08}[r][c][0.18]{$-0.08$}
\psfrag{-0.1}[r][c][0.18]{$-0.10$}
\psfrag{-16}[r][c][0.18]{$-16$}
\psfrag{-10}[r][b][0.18]{$-10$}
\psfrag{-1}[r][t][0.15]{$-1$}
\psfrag{-2}[r][t][0.15]{$-2$}
\psfrag{-3}[r][t][0.15]{$-3$}
\psfrag{-4}[r][t][0.15]{$-4$}
\psfrag{0}[r][c][0.18]{$0$}
%
%%x-axis
\psfrag{1}[c][b][0.18]{$1$}
\psfrag{5}[c][b][0.18]{$5$}
\psfrag{4}[c][b][0.18]{$4$}
\psfrag{6}[c][b][0.18]{$6$}
\psfrag{1.5}[c][b][0.18]{$1.5$}
\psfrag{5.5}[c][b][0.18]{$5.5$}
\psfrag{4.5}[c][b][0.18]{$4.5$}
\psfrag{3.5}[c][b][0.18]{$3.5$}
\psfrag{2.5}[c][b][0.18]{$2.5$}
\psfrag{2}[c][b][0.18]{$2$}
\psfrag{2.2}[c][b][0.18]{$2.2$}
\psfrag{2.4}[c][b][0.18]{$2.4$}
\psfrag{2.6}[c][b][0.18]{$2.6$}
\psfrag{2.8}[c][b][0.18]{$2.8$}
\psfrag{3}[c][b][0.18]{$3$}
}}
\caption{\ac{rs} and one-step \ac{rsb} approximation of the zero-temperature entropy for the system setup corresponding to the \ac{iid} \ac{rsb} curve in Fig. \ref{fig:1}. The figure confirms the better accuracy of the \ac{rsb} prediction.}
\label{fig:3}
\end{figure}

In order to study the accuracy of the one-step \ac{rsb} further, we invoke the common consistency test based on the zero-temperature entropy $\rm H^0$. In fact, considering the spin glass defined in \eqref{eq:6},  the distribution of the microstate at the zero temperature tends to an indicator function at the point which minimizes the Hamiltonian. Therefore, the entropy tends to zero. It has been, however, observed that in problems which \ac{rs} clearly fails, the zero-temperature entropy is also predicted wrongly in the sense that it does not tend to zero, but becomes negative. It has been further shown that for these cases the zero-temperature entropy under the RSB ans\"atze takes values much closer to zero. Fig. \ref{fig:3} shows the zero-temperature entropy of the corresponding spin glass under both the \ac{rs} and \ac{rsb} assumptions versus the compression rate. The system setup has been considered as in Fig. \ref{fig:1} and $\rm H^0$ has been determined at the tuning factors which minimize the one-step \ac{rsb} predicted \ac{mse}. As the compression rate grows, the \ac{rs} zero-temperature entropy drops down. The one-step \ac{rsb}, however, gives a better approximation for $\rm H^0$.

\bibliographystyle{IEEEtran}
\bibliography{ref}
%\begin{thebibliography}{1}
%\end{thebibliography}

\label{list:acronyms}
\begin{acronym}
\acro{iid}[i.i.d.]{independent and identically distributed}
\acro{pmf}[PMF]{Probability Mass Function}
\acro{cdf}[CDF]{Cumulative Distribution Function}
\acro{pdf}[PDF]{Probability Density Function}
\acro{rs}[RS]{Replica Symmetry}
\acro{1rsb}[1RSB]{One-Step Replica Symmetry Breaking}
\acro{brsb}[$b$RSB]{$b$-Steps Replica Symmetry Breaking}
\acro{rsb}[RSB]{Replica Symmetry Breaking}
\acro{mse}[MSE]{Mean Square Error}
\acro{map}[MAP]{Maximum-A-Posteriori}
\acro{rhs}[r.h.s.]{right hand side}
\acro{lhs}[l.h.s.]{left hand side}
\acro{wrt}[w.r.t.]{with respect to}
\end{acronym}

\end{document}